\documentclass{article}
\usepackage{arxiv}
\usepackage[utf8]{inputenc}
\usepackage[T1]{fontenc}
\usepackage{hyperref}
\usepackage{url}
\usepackage{booktabs}
\usepackage{amsfonts}
\usepackage{nicefrac}
\usepackage{microtype}
\usepackage{lipsum}
\usepackage{graphicx}
\usepackage[numbers]{natbib} 
\setcitestyle{notesep={; }} 
\usepackage{doi}
\usepackage{longtable}
\usepackage{booktabs}
\usepackage{multirow}
\usepackage{pifont}
\newcommand{\cmark}{\ding{51}}
\newcommand{\xmark}{\ding{55}}
\newcommand{\pmark}{\ding{108}}

\title{Unmasking Synthetic Realities in Generative AI: \\ A Comprehensive Review of Adversarially Robust Deepfake Detection Systems}

\author{ \href{https://orcid.org/0009-0008-2352-6845}{\includegraphics[scale=0.06]{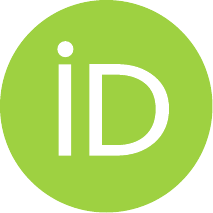}\hspace{1mm}Naseem Khan} \\
	Department of Computer Science\\
	Hamad bin Khalifa University\\
	Qatar \\
	\texttt{nakh12498@hbku.edu.qa} \\
	\And
	\href{https://orcid.org/0009-0000-2825-1547}{\includegraphics[scale=0.06]{orcid.pdf}\hspace{1mm}Tuan Nguyen} \\
	Qatar Computing Research Institute\\
	Hamad bin Khalifa University\\
	Qatar \\
	\texttt{ntuan@hbku.edu.qa} \\
	\And
	\href{https://orcid.org/0000-0003-4984-6093}{\includegraphics[scale=0.06]{orcid.pdf}\hspace{1mm}Amine Bermak} \\
	Department of Computer Science\\
	Hamad bin Khalifa University\\
	Qatar \\
	\texttt{abermak@hbku.edu.qa} \\
	\And
	\href{https://orcid.org/0000-0002-7660-9512}{\includegraphics[scale=0.06]{orcid.pdf}\hspace{1mm}Issa M. Khalil} \\
	Qatar Computing Research Institute\\
	Hamad bin Khalifa University\\
	Qatar \\
	\texttt{ikhalil@hbku.edu.qa} \\
}

\renewcommand{\shorttitle}{Unmasking Synthetic Realities in the GAI Era}

\hypersetup{
pdftitle={Unmasking Synthetic Realities in the GAI Era},
pdfsubject={q-bio.NC, q-bio.QM},
pdfauthor={David S.~Hippocampus, Elias D.~Striatum},
pdfkeywords={First keyword, Second keyword, More},
}

\begin{document}
\maketitle

\begin{abstract}

The rapid advancement of Generative Artificial Intelligence (GAI) has fueled the proliferation of deepfakes—synthetic media encompassing both fully generated content and subtly edited authentic material—posing profound challenges to digital security, misinformation mitigation, and identity preservation. This systematic review critically evaluates state-of-the-art deepfake detection methodologies, with an emphasis on reproducible, publicly available implementations to foster transparency and scientific validation. The analysis delineates two core paradigms: (1) the detection of fully synthetic media, leveraging statistical anomalies and hierarchical feature extraction, and (2) the localization of manipulated regions within authentic content, often employing multi-modal cues such as visual artifacts and temporal inconsistencies. These approaches, spanning uni-modal and multi-modal frameworks, demonstrate notable precision and adaptability in controlled settings, effectively identifying manipulations through advanced learning techniques and cross-modal fusion.

However, a comprehensive assessment reveals a pervasive limitation: the insufficient evaluation of adversarial robustness across both paradigms. Current methods, while proficient against known generative techniques, exhibit vulnerability to adversarial perturbations—subtle, intentional alterations designed to evade detection—undermining their reliability in real-world adversarial contexts. This gap highlights a critical disconnect between methodological development and the evolving threat landscape of GAI-driven attacks. To address this, we contribute a curated GitHub repository aggregating open-source implementations of the reviewed methods, enabling researchers to replicate, extend, and stress-test these approaches. Our findings emphasize the urgent need for future work to prioritize adversarial resilience, advocating for the design of scalable, modality-agnostic architectures capable of withstanding sophisticated manipulations. This review not only synthesizes the strengths and shortcomings of contemporary deepfake detection but also charts a path toward robust, trustworthy systems amid escalating digital threats. Link to github repository: \url{https://github.com/Magnet200/SOT_Deepfake_Detection_Mechanisms}
\end{abstract}


\section{Introduction}

Generative Artificial Intelligence (GAI) refers to the class of AI models designed to generate synthetic data that closely resembles real-world input. Initially developed to augment imbalanced datasets using techniques such as Generative Adversarial Networks (GANs), Variational Autoencoders (VAEs), and Autoregressive models, GAI has evolved into a powerful paradigm driving advancements across multiple domains \cite{sakirin2023survey, cao2023comprehensive}. Beyond its foundational role in data augmentation, GAI has revolutionized content creation, enabling human-computer collaboration (HCC) in areas such as art, music, literature, healthcare, and scientific research \cite{tan2023leveraging, liu2024collaboration}. Its widespread adoption has significantly impacted industries by improving efficiency, reducing costs, and fostering innovation \cite{ali2024constructing}. \par

Modern GAI models generate content across diverse modalities, including text, images, audio, video, and code. Prominent applications such as ChatGPT, Bard (Gemini), Midjourney, Copilot, DALL-E, and Synthesia demonstrate its broad utility \cite{gupta2024generative, Bu2023ResearchOD}. While these technologies offer substantial benefits, they also introduce ethical, security, and privacy concerns. The ability of generative models to produce highly realistic yet synthetic media raises concerns about misinformation, intellectual property rights, and privacy breaches. Consequently, research efforts increasingly focus on developing ethical and regulatory frameworks to ensure responsible deployment and forensic analysis of generative systems \cite{baldassarre2023social, dorigo2025artificial}. \par

Despite their transformative potential, GAI systems pose significant risks when misused. In education, the unrestricted use of generative models undermines academic integrity and critical thinking skills \cite{Simonsson1765190}. In business, inadequate regulatory controls expose organizations to security vulnerabilities and ethical dilemmas \cite{humphreys2024ai}. In healthcare, biased training data can lead to inaccurate AI-assisted diagnostics and decision-making \cite{chen2024generative}. Moreover, the propensity of GAI models to memorize and regenerate sensitive training data raises privacy concerns, particularly in journalism and legal contexts where confidential information must be safeguarded \cite{nishal2024envisioning}. These risks underscore the broader societal implications of GAI, including job displacement, socio-economic inequalities, and the potential weaponization of synthetic content for manipulation and disinformation campaigns \cite{popa2024critical}. \par

\begin{figure}[htbp]
    \centering
    \includegraphics[width=0.5\textwidth]{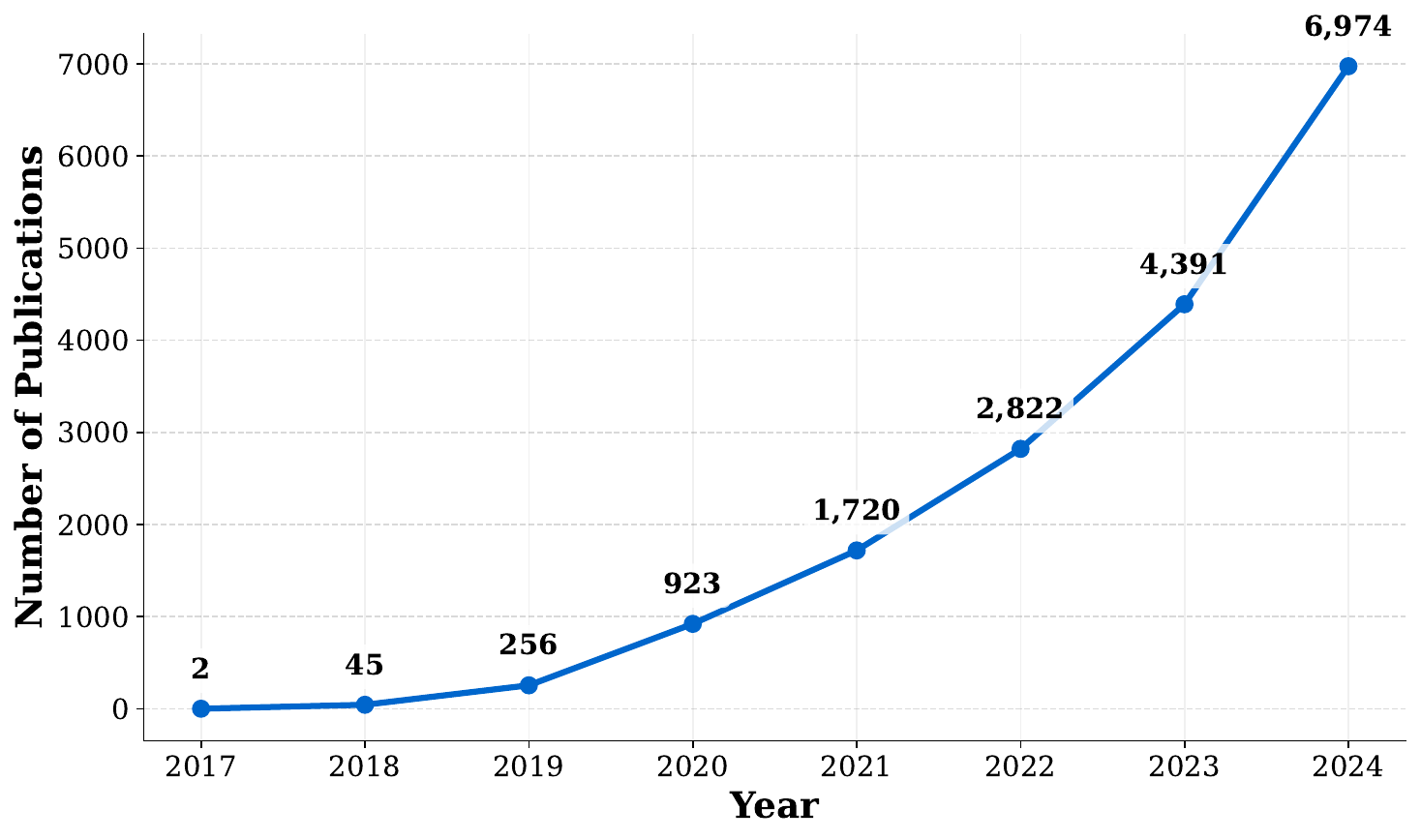}
    \caption{The graph illustrates the annual publication count in the field of DeepFakes. The data, obtained from dimensions.ai \cite{DimensionsAI}, highlights the development trend of Deepfakes detection from 2017 to 2024.}
    \label{fig:pub_trend}
\end{figure}

\begin{figure}[htbp]
    \centering
    \includegraphics[width=\textwidth]{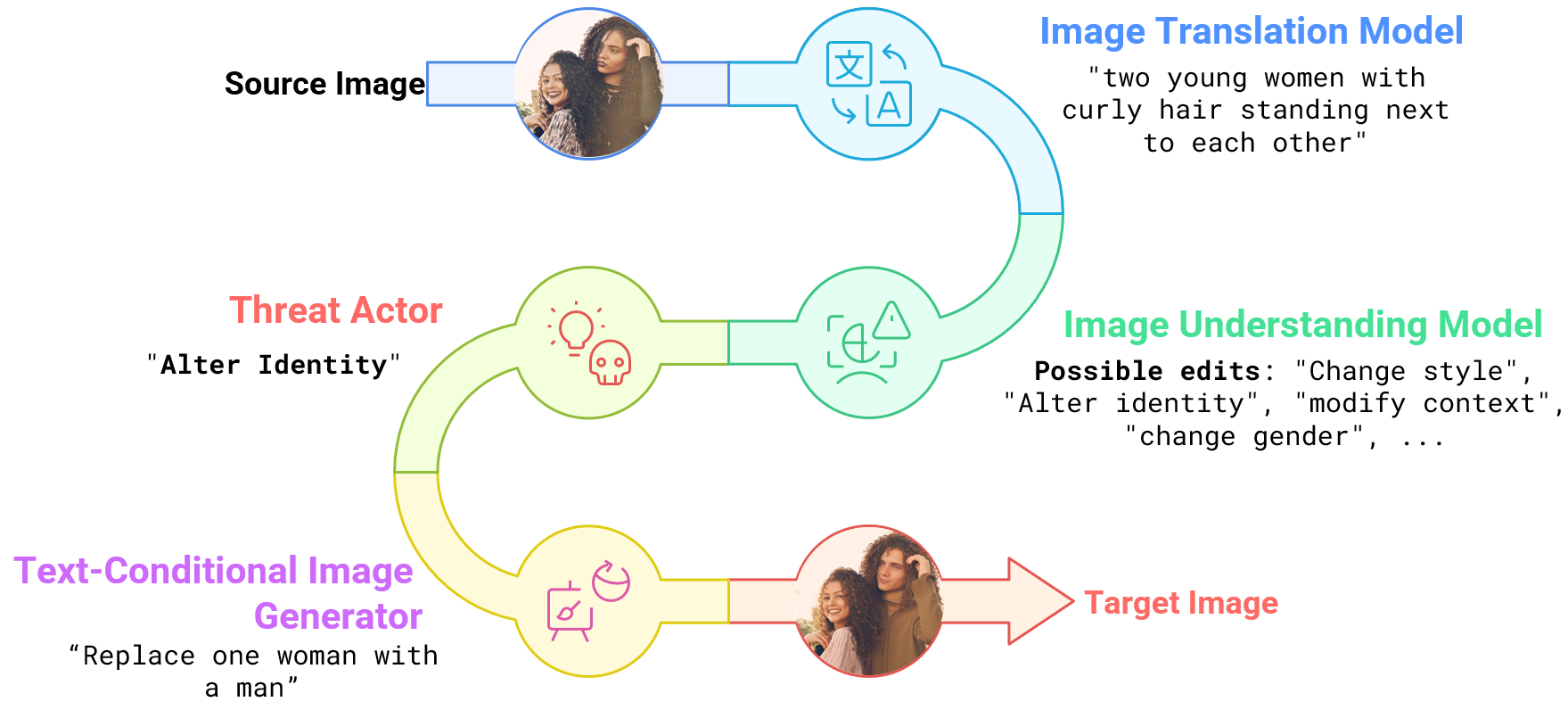}
    \caption{Illustration of a multi-stage pipeline in which a threat actor manipulates a source image using a text-conditional image generator, guided by identity-altering instructions, to produce a synthetic target image with modified personal attributes.}
    \label{fig:Identity_extortion}
\end{figure}

Among the most pressing challenges associated with GAI is the proliferation of synthetic media, commonly known as DeepFakes. DeepFakes leverage generative models to manipulate visual, auditory, and textual content, posing substantial threats to digital security, democratic stability, and public trust. As illustrated in Figure \ref{fig:pub_trend}, research interest in DeepFake detection has surged in response to the increasing sophistication and accessibility of these technologies \cite{DimensionsAI}. Malicious use cases include disinformation campaigns, identity fraud, extortion, and non-consensual explicit content generation, as exemplified in Figure \ref{fig:Identity_extortion}. These developments highlight the urgent need for robust forensic techniques capable of identifying and mitigating synthetic media threats. In particular, advancing adversarially robust detection mechanisms is crucial to ensuring the integrity and trustworthiness of digital content \cite{whyte2023beyond}. \par

This systematic review explores the current state of DeepFake detection methodologies, with a focus on uni-modal and multi-modal approaches, adversarial robustness, and cross-domain generalization. By critically analyzing existing frameworks, we aim to identify gaps in detection strategies and propose directions for future research toward more resilient and scalable solutions.

\section{Background and Technical Foundation}
\label{Background}
The evolution of deepfake technology is intrinsically linked to advancements in generative artificial intelligence (GAI). Since its emergence in 2017, deepfake technology has rapidly progressed from rudimentary face-swapping applications to sophisticated synthetic multimedia generation systems. This advancement is primarily driven by developments in generative models, notably Generative Adversarial Networks (GANs) \cite{goodfellow2014generative}, Diffusion Models (DMs) \cite{ho2020denoising}, and Variational Autoencoders (VAEs) \cite{pu2016variational}, which collectively underpin modern deepfake creation.

Early generative models relied on standalone architectures, such as basic encoder-decoder frameworks. However, the field has since shifted toward sophisticated conditional generation paradigms \cite{Chen2023TextimageGD}, enabling precise control over synthetic outputs. This shift has facilitated the integration of multiple modalities—visual, auditory, and textual—enhancing both the quality and versatility of generated media. For example, Conditional GANs \cite{mirza2014conditional, zhang2019image, reed2016generative} allow generators to produce tailored synthetic content based on specific input conditions, such as facial landmarks or text prompts. Similarly, Diffusion Models \cite{zhang2023adding, saharia2022photorealistic, poole2022dreamfusion} employ iterative denoising processes to create high-fidelity synthetic media guided by user-defined constraints, offering a robust alternative to GAN-based approaches.

Recent innovations in multi-modal learning have further elevated synthetic media generation, particularly in text-to-image (T2I) and text-to-video (T2V) domains. Models like DALL-E \cite{ramesh2021zero} exemplify this capability by generating photorealistic images from textual descriptions, while T2V advancements leverage pre-trained visual representations to produce coherent video sequences without requiring paired text-video training data \cite{ho2022video}. These developments are supported by spatial-temporal architectures that enhance motion modeling and resolution, resulting in highly realistic video content \cite{ho2022video}. Such progress has significantly expanded the scope of deepfake applications across modalities.

\begin{table}[t]
\caption{Comparative analysis of prior systematic reviews on deepfake detection, highlighting their modality coverage, detection paradigm and evaluation focus.}
\label{tab:comparison}
\footnotesize
\begin{tabular}{p{3.5cm}cccccc|c}
\toprule
\textbf{Coverage Aspects} & \textbf{\cite{le2024sok}} & \textbf{\cite{Kaur2024DeepfakeVD}} & \textbf{\cite{Yi2023AudioDD, li2025survey}} & \textbf{\cite{alnabhan2024fake}} & \textbf{\cite{kumar2025advances}} & \textbf{\cite{Liu2024EvolvingFS}} & \textbf{Our Study} \\
\midrule
\multicolumn{8}{l}{\textit{Detection Modalities}} \\
\midrule
Image-based & \cmark & \pmark & \xmark & \xmark & \pmark & \cmark &  \cmark \\
Video-based & \cmark & \cmark & \xmark & \xmark & \pmark & \cmark &  \cmark \\
Audio-based & \xmark & \xmark & \cmark & \xmark & \pmark & \xmark &  \cmark \\
Text-based & \xmark & \xmark & \xmark & \cmark & \pmark & \xmark &  \cmark \\
Multi-modal & \xmark & \xmark & \xmark & \xmark & \cmark & \cmark &  \cmark \\
\midrule
\multicolumn{8}{l}{\textit{Detection Paradigms}} \\
\midrule
Fully Synthetic & \cmark & \cmark & \cmark & \cmark & \cmark & \cmark &  \cmark \\
Edited Region Localization & \pmark & \xmark & \xmark & \xmark & \xmark & \pmark & \cmark \\
\midrule
\multicolumn{8}{l}{\textit{Evaluation Metrics}} \\
\midrule
Cross-dataset Generalization & \cmark & \pmark & \pmark & \pmark & \pmark & \cmark &  \cmark \\
Natural Perturbations & \pmark & \xmark & \pmark & \xmark & \xmark & \pmark &  \cmark \\
Adversarial Robustness & \xmark & \xmark & \cmark & \xmark & \pmark & \xmark &  \cmark \\
\bottomrule
\multicolumn{8}{p{14cm}}{\scriptsize \cmark~Comprehensive coverage, \pmark~Partial coverage, \xmark~Limited/No coverage} \\
\end{tabular}
\end{table}

In the visual domain, deepfakes benefit from conditional image synthesis and high-resolution video generation techniques. Audio deepfakes, driven by models such as WaveNet \cite{oord2016wavenet} and Tacotron \cite{wang2017tacotron}, utilize generative architectures trained on speech data to produce realistic synthetic voices, often synchronized with visual elements for enhanced authenticity. Text-based synthetic content, powered by large language models like GPT \cite{radford2019language}, generates human-like narratives that can complement other modalities. The convergence of these advances has given rise to multi-modal deepfakes, where synchronized audio, visual, and textual components create highly convincing synthetic media. High-profile cases, such as political misinformation campaigns and identity spoofing \cite{westerlund2019emergence}, highlight the growing realism and accessibility of this technology.

This increased sophistication, however, presents formidable challenges for detection systems. Traditional methods, which often exploit modality-specific artifacts—such as those in GAN-generated images \cite{wang2019cnn}—struggle against modern deepfakes. Diffusion-based models, for instance, produce fewer detectable traces, complicating forensic analysis \cite{wang2023diffusion}. Moreover, these generative frameworks can be exploited through adversarial techniques, crafting deepfakes designed to evade detection algorithms. Subtle perturbations, as explored in adversarial attack research \cite{hussain2021adversarial}, can deceive classifiers, exposing vulnerabilities in existing detection frameworks, especially in real-world adversarial settings.

The seamless integration of multi-modal cues and the emergence of adversarially crafted threats underscore the urgent need for advanced detection strategies. These must address not only the realism of modern deepfakes but also their intentional manipulations across modalities. The remainder of this review critically evaluates state-of-the-art uni-modal and multi-modal deepfake detection approaches, with a focus on enhancing adversarial robustness to counter these evolving synthetic challenges.

\section{Systematic Review Methodology}
\label{Systematic Review Methodology}
To systematically assess advancements in deepfake detection, we conducted a structured literature review, emphasizing model robustness, transferability, and multi-modal approaches. Traditional detection methods rely on publicly available datasets, limiting their effectiveness against novel generative architectures. The rapid evolution of deepfake generation, particularly with techniques such as Low-Rank Adaptation (LoRA) \cite{hu2022lora}, has increased the diversity of synthetic content, making exhaustive model-specific training impractical. This review, therefore, examines the generalization capability of detection models across datasets and their adaptability to emerging threats.

\subsection{Search Strategy and Inclusion Criteria}
\label{subsec:search_strategy}

Our review follows a systematic approach for selecting relevant literature. We queried the following databases:
\begin{itemize}
    \item \textbf{Databases}: Dimensions.ai, Semantic Scholar, IEEE Xplore, ACM Digital Library, and arXiv.
    \item \textbf{Search Terms}: "Deepfake detection," "Image/Audio/Video/Text-based deepfake detection," "adversarial robustness in AI forensics," "uni-modal deepfake detection," "multi-modal deepfake detection," "cross-domain generalization in synthetic media detection."
    \item \textbf{Timeframe}: Studies published from January 2023 to early 2025 were prioritized to ensure coverage of the latest advancements.
    \item \textbf{Inclusion Criteria}: Peer-reviewed journal and conference papers focusing on image, video, audio, text-based, and multi-modal deepfake detection.
    \item \textbf{Exclusion Criteria}: Papers without empirical results, non-English publications, and theoretical discussions without publicly available implementation.
\end{itemize}
Each study underwent a systematic quality assessment evaluating: (1) relevance to deepfake analysis domains, (2) availability of public implementation repositories, (3) experimental validation on contemporary benchmark datasets, and (4) methodological innovation relative to existing approaches—ensuring our review encompasses technically sound, reproducible, and state-of-the-art contributions.

\subsection{Scope of Existing Systematic Reviews}
\label{subsec:existing_reviews}

The field of deepfake detection has seen extensive research, leading to multiple systematic reviews. However, existing surveys often focus on specific modalities rather than comprehensively addressing deepfake detection across different forms of synthetic media. Table~\ref{tab:comparison} provides an overview of prior surveys, highlighting their scope, strengths, and limitations.

Most prior reviews target image and video-based detection, with notable contributions focusing on facial deepfake analysis and fully synthetic content identification \cite{le2024sok, Kaur2024DeepfakeVD}. While these studies provide a strong foundation, they typically exclude audio and text-based manipulations, limiting their applicability in multi-modal contexts. Similarly, surveys on audio deepfake detection focus on speech synthesis and voice conversion techniques but do not address cross-modal threats that combine visual and auditory elements \cite{Yi2023AudioDD, li2025survey}.  

Text-based deepfake detection, particularly in misinformation and fake news detection, has been explored in recent studies \cite{alnabhan2024fake}. However, these works primarily analyze linguistic patterns and do not integrate insights from other modalities. Some broader surveys attempt to cover multiple modalities \cite{kumar2025advances, Liu2024EvolvingFS}, yet they lack a comprehensive adversarial robustness evaluation and fail to systematically assess cross-dataset generalization and real-world transferability.  

While these reviews have significantly advanced the field, none provide a unified framework that integrates deepfake detection across all media formats, including image, video, audio, text, and multi-modal systems. Additionally, most prior studies focus on conventional detection approaches without addressing emerging challenges such as adversarial attacks, generative fine-tuning, and imperceptible content manipulations.

\subsection{Contributions of This Review}
\label{subsec:contributions}

This systematic review provides a comprehensive synthesis of deepfake detection methodologies, encompassing all primary modalities—image, video, audio, text, and multi-modal systems. It bridges critical gaps in the literature by integrating uni-modal and multi-modal approaches while offering a structured evaluation of their resilience to adversarial threats and their adaptability across diverse synthetic media contexts.

A key contribution lies in the classification of detection strategies into two distinct paradigms: (i) the identification of entirely synthetic media and (ii) the localization of manipulated regions within authentic content. The latter, an emerging and underexplored domain, demands sophisticated forensic techniques and adaptive analytical methods, which this review systematically explores.

Additionally, this study delivers an in-depth examination of cross-domain generalization, a crucial attribute for ensuring detection systems remain effective against evolving generative technologies. By assessing the transferability of detection approaches across varied synthetic media types, this review establishes a foundation for understanding their robustness in dynamic, real-world scenarios.

Adversarial robustness is a cornerstone of this analysis, with a critical evaluation of how contemporary detection systems withstand real-world perturbations and adversarial modifications. This includes a discussion of countermeasures—such as adversarial training and feature-space regularization—designed to bolster resilience against advanced manipulation techniques, without reliance on specific attack frameworks.

\begin{figure}[htbp]
    \centering
    \includegraphics[width=\textwidth]{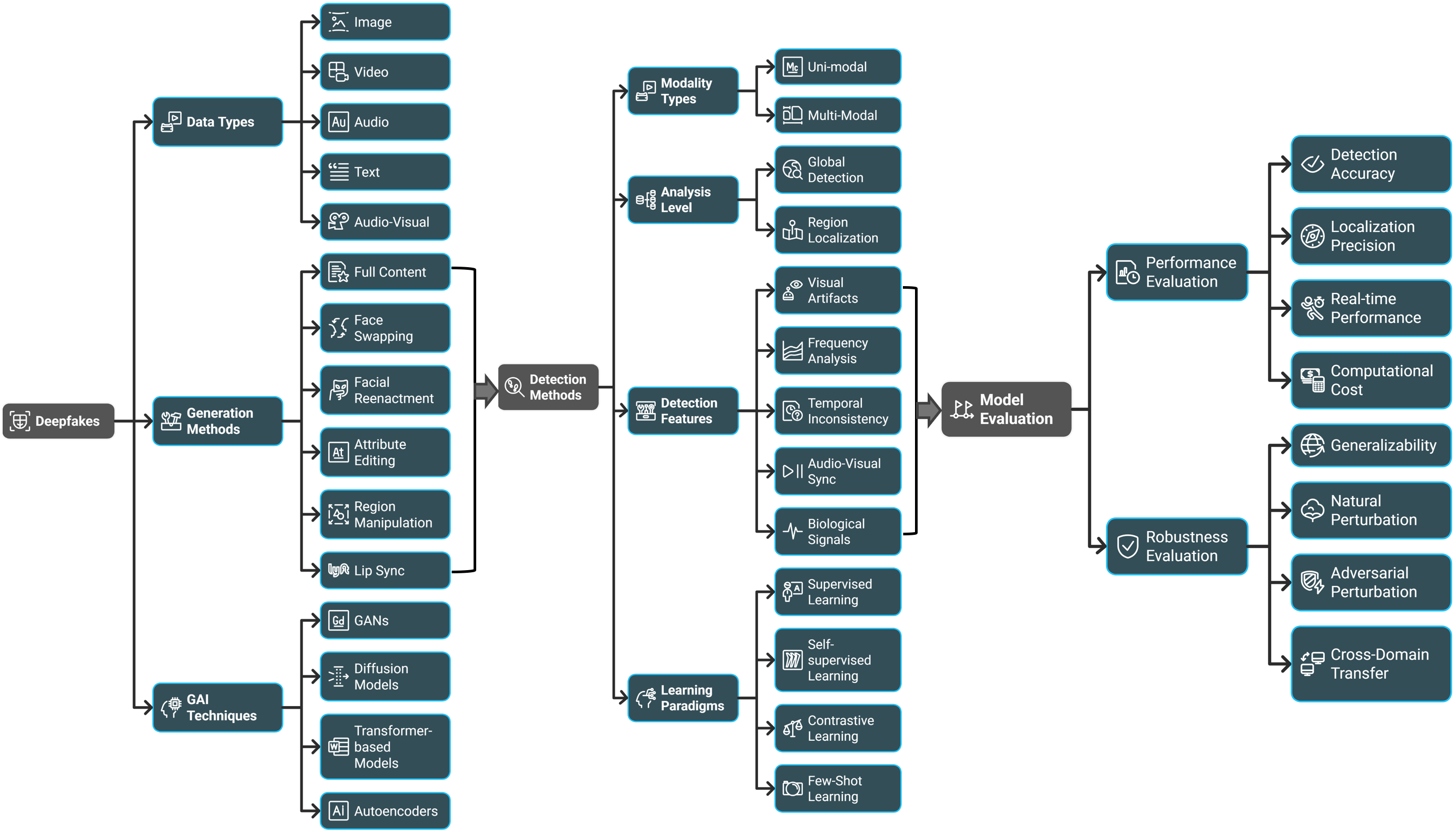}
    \caption{The broad taxonomy of Deepfake generation and detection strategies.}
    \label{fig:Deepfake_taxonomy}
\end{figure}

To uphold scientific integrity and support practical implementation, this review emphasizes studies with publicly accessible implementations, fostering reproducibility and enabling comparative assessments. A consistent evaluative framework is applied across all methodologies, scrutinizing their strengths, limitations, and applicability to operational contexts.

Finally, this review delineates key research gaps and proposes future directions, including the incorporation of explainable AI for enhanced forensic transparency, the advancement of real-time detection capabilities, and the exploration of self-supervised learning to improve generalization and robustness. By consolidating current knowledge and outlining a trajectory for future inquiry, this study serves as a vital resource for developing resilient detection systems to counter the growing sophistication of generative AI-driven synthetic media.

\subsection{Conclusion and Transition to Detection Methodologies}
\label{subsec:conclusion}

This review consolidates prior research on deepfake detection, emphasizing its strengths and gaps. Unlike previous surveys, it systematically integrates multi-modal detection, adversarial robustness analysis, and cross-dataset generalization. The following section explores state-of-the-art detection methodologies, evaluating their efficacy in combating the increasing sophistication of GAI.

\section{Deepfake Detection}
\label{Deepfake Detection}
Deepfake detection methods can be broadly categorized into \textit{Uni-modal} and \textit{Multi-modal} approaches, depending on whether they analyze a single data type or integrate multiple modalities for classification. Uni-modal methods focus on domain-specific artifacts within images, audio, or text, leveraging spatial inconsistencies, frequency distortions, or linguistic anomalies to detect manipulations. In contrast, multi-modal approaches enhance robustness by combining complementary information across modalities, such as synchronizing facial expressions with speech in audio-visual deepfake detection. Advanced fusion techniques, including attention mechanisms and contrastive learning, further improve cross-modal consistency analysis, enabling more reliable detection of synthetic content. The broad taxonomy of Deepfake generation and detection approaches depicted in Figure \ref{fig:Deepfake_taxonomy}.

\subsection{Feature Learning Strategies in Deepfake Detection}
\label{Feature_learning_strategies}

Beyond modal categorization, deepfake detection techniques can be classified based on their \textit{feature extraction and learning mechanisms}. Traditional approaches primarily focus on spatial pattern learning, frequency spectrum analysis, and temporal consistency modeling. Spatial-based methods detect visual artifacts, inconsistencies in texture, or pixel-level anomalies introduced during synthesis \cite{tolosana2020deepfakes}. Frequency-based techniques exploit traces left by generative models in the frequency domain, leveraging Fourier or wavelet transforms to capture imperceptible patterns \cite{frank2020leveraging}. Temporal consistency learning extends these methods by analyzing motion coherence across frames to detect artifacts in synthesized videos \cite{sabir2019recurrent}. Textual-based methods identify linguistic anomalies or contextual inconsistencies in AI-generated text, such as fake news, using deep learning models like BERT \cite{Chong2023BotOH}.

Recent advancements have explored physiological signal analysis, such as heart rate variations (rPPG) and facial micro-expressions, to differentiate real and fake content \cite{ciftci2020fakecatcher}. Another emerging approach involves detecting semantic and contextual inconsistencies, focusing on unnatural facial expressions, lip-sync mismatches, or logical errors in AI-generated text \cite{agarwal2020detecting}. To enhance robustness, adversarial perturbation analysis is employed, evaluating how detection models respond to adversarial attacks or subtle perturbations designed to bypass classifiers \cite{hussain2021adversarial}. Furthermore, self-supervised and contrastive learning methods have gained attention for their ability to generalize across different generative models by learning invariant feature representations \cite{li2021frequency}. Explainability-driven strategies, leveraging techniques like Grad-CAM, SHAP, TSNE, or textual description, also play a role in improving interpretability by highlighting manipulated regions in images or videos \cite{malolan2020explainable}.

These diverse learning strategies collectively contribute to improving deepfake detection, enhancing model generalizability, and addressing evolving generative AI techniques. To systematically assess the strengths and limitations of these detection strategies, the following sections provide an in-depth review of \textit{Uni-modal detection approaches}, evaluating their \textit{model architectures, benchmark datasets used, performance metrics in both natural and adversarial adaptability, and their generalization ability against unseen generative models}.

\subsubsection{Uni-modal Deepfake Detection}

\textbf{Image deepfakes:} 
Image-based deepfake detection has been a primary focus due to the proliferation of generative models producing realistic synthetic images, ranging from fully synthetic faces to subtle manipulations. Early methods relied on \textit{spatial pattern learning} using convolutional neural networks (CNNs) to detect pixel-level inconsistencies \cite{Wang2019FakeSpotterAS, tan2023learning}. However, their limited adaptability to diverse generative sources prompted a shift to \textit{frequency-domain analysis}, leveraging statistical traces to enhance generalization \cite{tan2024frequency, Jeong2022FrePGANRD, doloriel2024frequency, pontorno2024exploitation, tian2023frequency, liu2024detection, lu2024towards, coccomini2024deepfake}. Recent approaches integrate \textit{self-supervised} and \textit{contrastive learning} to improve feature invariance and robustness \cite{baru2025wavelet, ojha2023towards, tang2025towards, liu2024forgery, xu2025recent}, while lightweight architectures prioritize efficiency \cite{Lanzino2024FasterTL, Lim2024DistilDIREAS}. Robustness against adversarial perturbations is addressed through watermarking and adaptive frameworks \cite{Wu2023SepMarkDS, Hsu2024AIassistedDD, lukas2023ptw, wang2024lampmark, qiao2024scalable}, with explainability-driven strategies enhancing interpretability \cite{gowrisankar2024adversarial, Pontorno2024DeepFeatureXND, Aghasanli2023InterpretablethroughprototypesDD}.

\begingroup
\scriptsize 
\setlength{\tabcolsep}{3pt} 
\renewcommand{\arraystretch}{1.3} 

\begin{longtable}{@{}p{1.0cm}p{2.2cm}p{2.5cm}p{3.8cm}p{4.5cm}@{}} 
\caption{Uni-modal Deepfake Detection: Synthesis of Mechanisms and Insights} \label{tab:unimodal_synthesis} \\
\toprule
\scriptsize 
\textbf{Modality} & \textbf{Approach} & \textbf{References} & \textbf{Strength} & \textbf{Challenge} \\
\midrule
\endfirsthead

\multicolumn{5}{c}{\tablename\ \thetable{} -- Continued from previous page} \\
\toprule
\scriptsize
\textbf{Modality} & \textbf{Approach} & \textbf{References} & \textbf{Strength} & \textbf{Challenge} \\
\midrule
\endhead

\midrule
\multicolumn{5}{r}{Continued on next page} \\
\endfoot

\bottomrule
\multicolumn{5}{p{13.6cm}}{\scriptsize \textbf{Notes:} FF++ = FaceForensics++, DFDC = DeepFake Detection Challenge.} \\
\endlastfoot

\multicolumn{5}{l}{\textbf{Image-based Detection}} \\
\midrule
\multicolumn{5}{l}{\parbox{14.0cm}{\raggedright\tiny Datasets: CelebA, FFHQ, ForenSynths, LSUN, DiffusionForensics, CNNSpot, ProGAN, COCO, ImageNet, LAION, GANGen-Detection, UnivFakeDetect, MSCOCO}} \\
\midrule
Image & Frequency-Based & \tiny{\cite{Tan2024DataIndependentOA, tan2024frequency, Jeong2022FrePGANRD, Doloriel2024FrequencyMF, Pontorno2024OnTE, Lu2024TowardsTD, Coccomini2024DeepfakeDW, Liu2024DetectionOD, li2025optimized}} & Uncovers subtle synthesis artifacts & Limited by evolving generative techniques \\
Image & Spatial-Based & \tiny{\cite{Wang2019FakeSpotterAS, tan2023learning, tan2024rethinking, Pontorno2024DeepFeatureXND, zheng2025breaking, wang2023dire, Guarnera2024MasteringDD, Tan2024DataIndependentOA}} & Captures pixel-level inconsistencies & Sensitive to image degradation \\
Image & Adaptive Learning & \tiny{\cite{tian2023frequency, liu2024forgery, ojha2023towards, Aghasanli2023InterpretablethroughprototypesDD, abdullah2024analysis, Lim2024DistilDIREAS, tang2025towards, baru2025wavelet, subudhi2024adaptive}} & Adapts to diverse generative models & Requires extensive training data \\
Image & Robust Learning & \tiny{\cite{gowrisankar2024adversarial, Lanzino2024FasterTL, dhanakshirur2024herd, chen2024masked, subudhi2024adaptive}} & Enhances stability under disruptions & High computational complexity \\
Image & Watermarking & \tiny{\cite{Wu2023SepMarkDS, Hsu2024AIassistedDD, lukas2023ptw, wang2024lampmark, qiao2024scalable}} & Embeds detectable authenticity cues & Vulnerable to sophisticated attacks \\

\multicolumn{5}{l}{\textbf{Video-based Detection}} \\
\midrule
\multicolumn{5}{l}{\parbox{14.0cm}{\raggedright\tiny Datasets: FF++, Celeb-DF, DFDC, WildDeepfake, FakeAVCeleb, DeeperForensics, ForgeryNet, DFD, Seq-DeepFake, KODF-LS, LSR+W2L, DF40}} \\
\midrule
Video & Frame-Based & \tiny{\cite{zhu2023face, xu2024rlgc, Lin2024PreservingFG, she2024using, Nguyen2024LAANetLA, shao2022detecting, zhang2024samif, zhu2021face}} & Leverages static visual cues & Misses temporal inconsistencies \\
Video & Temporal-Based & \tiny{\cite{zhang2024temporal, song2024quality, hasanaath2025fsbi, Bah2024EnhancedDD, Datta2024ExposingLD, xu2024learning, shao2025robust, chen2024demamba, he2024lip}} & Detects motion-based anomalies & Dependent on video quality \\
Video & Temporal+Graph & \tiny{\cite{yang2023masked, xie2025grdt, sun2023contrastive, ba2024exposing}} & Models relational dynamics & Complex model training \\
Video & Adaptive Learning & \tiny{\cite{cai2023marlin, larue2023seeable, dong2023implicit, kong2024open, Tan2024C2PCLIPIC, Lin2024PreservingFG, peng2023deepfidelity}} & Handles varied forgery types & Limited by dataset diversity \\
Video & Robust Learning & \tiny{\cite{shao2025deepfake, gong2025robust, peng2023deepfidelity, tsigos2025improving, Wodajo2024ImprovedDV}} & Improves reliability in real conditions & Resource-intensive processing \\
Video & Hybrid/Advanced & \tiny{\cite{kaddar2024deepfake, siddiqui2025enhanced, mahmud2023unmasking}} & Combines local and global features & Sensitive to low-quality inputs \\

\multicolumn{5}{l}{\textbf{Audio-based Detection}} \\
\midrule
\multicolumn{5}{l}{\parbox{13.0cm}{\raggedright\tiny Datasets: ASVspoof 2015, ASVspoof 2019, ASVspoof 2021, In-the-Wild, FakeAVCeleb, CVoiceFake, SONICS, WaveFake, EVDA, CLEAR, VSA, GRID, CD-ADD}} \\
\midrule
Audio & Artifact-Focused & \tiny{\cite{Shih2024DoesAD, Zhang2023WhatTR, chen2024region, doan2023bts, klein2024source}} & Identifies synthesis imperfections & Fails with high-fidelity fakes \\
Audio & Temporal Modeling & \tiny{\cite{Shin2023HMCONFORMERAC, Chen2024RawBMambaEB, rahman2024sonics}} & Captures sequential patterns & Vulnerable to noise interference \\
Audio & Adaptive Learning & \tiny{\cite{Xie2024DomainGV, yang2024robust, zhu2025slim, zhang2024audio, oiso2024prompt}} & Adapts to new synthesis methods & Relies on broad data coverage \\
Audio & Robust Learning & \tiny{\cite{Lee2025DualChannelDA, wu2024clad, Li2024CrossDomainAD, xie2024does, Li2024SafeEarCP, Wani2024ABCCapsNetAB, combei2024wavlm}} & Ensures stability across environments & Struggles with atypical audio \\

\multicolumn{5}{l}{\textbf{Text-based Detection}} \\
\midrule
\multicolumn{5}{l}{\parbox{13.0cm}{\raggedright\tiny Datasets: ISOT, TweepFake, OpenLLMText, PHEME, FA-KES, WebText, Ch-9, RealNews, Enhanced TweepFake, SynSciPass}} \\
\midrule
Text & Linguistic-Based & \tiny{\cite{tembhurne2022mc, amutha2024detection, wang2024spectral}} & Exploits syntactic anomalies & Limited by style variations \\
Text & Transformer-Based & \tiny{\cite{Chong2023BotOH, Uchendu2023TopFormerTA, lee2024enhancing, alnabhan2024bertguard, li-etal-2024-mage}} & Leverages contextual understanding & High computational demands \\
Text & Hybrid/Advanced & \tiny{\cite{pu2023deepfake}} & Integrates multiple features & Challenged by evolving LLMs \\

\end{longtable}
\endgroup

New advancements include \textit{Data-Independent Operator (DIO)} \cite{Tan2024DataIndependentOA}, a training-free method using handcrafted filters for artifact extraction, and \textit{Neighboring Pixel Relationships (NPR)} \cite{tan2024rethinking}, which captures local pixel correlations for source-invariant detection. \textit{Diffusion Reconstruction Error (DIRE)} \cite{wang2023dire} and its distilled variant \cite{Lim2024DistilDIREAS} leverage reconstruction errors to detect diffusion-generated images, while \textit{hierarchical frameworks} \cite{Guarnera2024MasteringDD} classify images across multiple levels. \textit{Subudhi et al.} \cite{subudhi2024adaptive} introduce a meta-learning approach for adaptability, and \textit{Abdullah et al.} \cite{abdullah2024analysis} propose ensemble methods with content-agnostic features to counter fine-tuned generative models. \textit{Chen et al.} \cite{chen2024masked} augment datasets with masked diffusion models, and \textit{Li et al.} \cite{li2025optimized} optimize spatial-frequency collaboration for IoT security. Challenges persist in generalizing to unseen models and resisting sophisticated adversarial attacks.

\textbf{Video deepfakes:} 
Video-based detection targets temporal inconsistencies in manipulated sequences. Early frame-based methods detected spatial artifacts \cite{shao2022detecting, hasanaath2025fsbi, Bah2024EnhancedDD, xu2024rlgc}, but overlooked inter-frame dynamics. \textit{Temporal consistency modeling} using 3D CNNs and transformers captures lip-sync mismatches and motion irregularities \cite{Datta2024ExposingLD, zhang2024temporal, xu2024learning}, while \textit{graph-based modeling} identifies relational anomalies \cite{yang2023masked, she2024using, xie2025grdt}. Generalization is enhanced through \textit{self-supervised} and \textit{contrastive learning} \cite{cai2023marlin, sun2023contrastive, larue2023seeable}, and pre-trained models with adapters \cite{shao2025deepfake, Tan2024C2PCLIPIC, kong2024open}. Robustness is improved via hybrid architectures \cite{kaddar2024deepfake, Wodajo2024ImprovedDV, gong2025robust}, with interpretability addressed through explainable techniques \cite{mahmud2023unmasking, tsigos2025improving, siddiqui2025enhanced}.

Recent contributions include \textit{Lin et al.} \cite{Lin2024PreservingFG}, ensuring fairness across demographics, and \textit{Zhu et al.} \cite{zhu2021face, zhu2023face}, decomposing frames into 3D components. \textit{SeqFakeFormer++} \cite{shao2025robust} detects sequential manipulations, while \textit{Song et al.} \cite{song2024quality} use quality-centric enhancements. \textit{Peng et al.} \cite{peng2023deepfidelity} assess perceptual fidelity, and \textit{Ba et al.} \cite{ba2024exposing} employ information bottleneck principles. \textit{Dong et al.} \cite{dong2023implicit} mitigate identity leakage, and \textit{Nguyen et al.} \cite{Nguyen2024LAANetLA} focus on localized artifacts. Challenges remain in handling low-quality videos and unseen forgeries.

\textbf{Audio Deepfakes:} 
Audio detection targets synthetic speech and songs, evolving from artifact-based methods \cite{Shih2024DoesAD} to robust techniques leveraging \textit{non-verbal cues} like breathing \cite{doan2023bts, zhu2025slim} and \textit{temporal dependencies} via transformers \cite{Chen2024RawBMambaEB, rahman2024sonics}. Generalization is improved through continual learning and domain generalization \cite{Zhang2023WhatTR, chen2024region, Xie2024DomainGV, Li2024CrossDomainAD, zhang2024audio}, while robustness to perturbations employs augmentation \cite{Lee2025DualChannelDA, wu2024clad, Wani2024ABCCapsNetAB}. New methods include \textit{HM-Conformer} \cite{Shin2023HMCONFORMERAC} for hierarchical feature extraction, \textit{TDVSA-Net} \cite{he2024lip} for lip-based authentication, and \textit{DeMamba} \cite{chen2024demamba} for spatio-temporal analysis. \textit{Yang et al.} \cite{yang2024robust} fuse multi-view features, and \textit{Oiso et al.} \cite{oiso2024prompt} optimize prompt tuning. \textit{SafeEar} \cite{Li2024SafeEarCP} ensures privacy, while \textit{Klein et al.} \cite{klein2024source} trace sources. \textit{CtrSVDD} \cite{zang24_interspeech} targets singing voices, and \textit{Weizman et al.} \cite{weizman2024tandem} enhance ASV systems. \textit{Xie et al.} \cite{xie2024does} address ALM-based audio, and \textit{Combei et al.} \cite{combei2024wavlm} use ensemble learning. Adversarial robustness and cross-domain performance remain underexplored.

\textbf{Text Deepfakes:} 
Text detection targets synthetic content like fake news, advancing from \textit{linguistic feature extraction} \cite{tembhurne2022mc, Chong2023BotOH} to \textit{contextual analysis} \cite{amutha2024detection, Uchendu2023TopFormerTA} and \textit{domain-specific detection} \cite{alnabhan2024bertguard, wang2024spectral}. Generalization employs multi-task and zero-shot learning \cite{wang2024spectral, pu2023deepfake}, with robustness enhanced by adversarially fine-tuned models \cite{lee2024enhancing, pu2023deepfake, Uchendu2023TopFormerTA}. New methods like \textit{Mc-DNN} \cite{tembhurne2022mc} process multi-channel text, \textit{OBi-LSTM-CNN} \cite{amutha2024detection} optimize rumor detection, and \textit{MAGE} \cite{li-etal-2024-mage} tackle diverse LLMs. Challenges include adapting to evolving LLMs and resisting adversarial text modifications.

\textbf{Conclusion:} 
Uni-modal deepfake detection has advanced considerably across modalities. Image-based methods leverage frequency analysis and self-supervised learning, video detection emphasizes temporal and graph-based modeling, audio techniques exploit non-verbal cues and transformers, and text detection refines contextual and domain-specific approaches. Despite progress, challenges in generalization to unseen sources, robustness against adversarial attacks, and cross-modal adaptability underscore the need for continued research.

\subsubsection{Multi-Modal Deepfake Detection Mechanisms}
The evolution of generative AI has escalated deepfake synthesis beyond uni-modal alterations—such as facial manipulations in static images or video frames—to intricate multi-modal fabrications integrating visual, auditory, and textual elements. Uni-modal detection strategies, while proficient in isolating modality-specific anomalies (e.g., pixel-level distortions or spectral irregularities), exhibit limited efficacy against cross-modal inconsistencies inherent in advanced generative systems, such as text-driven image synthesis or audio-visual misalignment. Multi-modal detection mechanisms address this deficiency by employing integrated learning frameworks—contrastive representation, modality fusion, and inconsistency modeling—to discern interdependencies and deviations across heterogeneous data streams, demonstrating superior detection accuracy compared to uni-modal counterparts. Illustrative examples include the identification of lip-audio desynchronization \cite{zhou2021joint, liu2024lips, Shahzad2023AVLipSyncLA} and text-visual incongruities \cite{Santosh2024RobustCD, Chang2023AntifakePromptPV}, which evade uni-modal scrutiny.

\begingroup
\scriptsize 
\setlength{\tabcolsep}{3pt} 
\renewcommand{\arraystretch}{1.2} 

\begin{table}[ht]
\centering
\caption{Overview of Multi-Modal Deepfake Detection Mechanisms} \label{tab:multimodal_overview}
\begin{tabular}{@{}p{3.0cm}p{3.3cm}p{2.5cm}p{6.3cm}@{}} 
\toprule
\textbf{Category} & \textbf{References} & \textbf{Modalities} & \textbf{Strengths and Challenges} \\
\midrule
\textbf{VLM} & \tiny{\cite{Santosh2024RobustCD, on-learning-multi-modal-forgery-representation-for-diffusion-generated-video-detection, Laiti2024ConditionedPF, li2024fakebench, Chang2023AntifakePromptPV, Zhang2024CommonSR, lai2024gm, foteinopoulou2024hitchhiker, Shao2023DetectingAG, Khan2024CLIPpingTD, sha2023fake, cozzolino2023raising, Jia2024CanCD, shah2024feature, keita2025bi, keita2024harnessing}} & Image+Text & Interpretable detection, generalizable across generators; limited by prompt dependency, high-fidelity fakes \\
\textbf{AV Sync} & \tiny{\cite{zhou2021joint, liu2024lips, Shahzad2023AVLipSyncLA, zou2024cross, liang2024speechforensics, Feng2023SelfSupervisedVF, zhang2024joint, Cozzolino2022AudioVisualPD, koutlis2024dimodif, li2024zero, mo2024unveiling}} & Audio+Video & Robust to natural perturbations, sync-focused; struggles with non-frontal faces, unsynchronized inputs \\
\textbf{MV Fusion} & \tiny{\cite{wu2024audio, yoon2024triple, Guo2024AVSecureAA, kharel2023df, oorloff2024avff, Katamneni2023MISAVoiDDMI, bekheet2024development, Yang2023AVoiDDFAJ, 10609529, lee2024multi, Liu2024MCLMC, pellicer2024pudd}} & Aud+Vid+Txt & Broad modality coverage, strong transferability; computationally intensive, alignment issues \\
\textbf{SS Learning} & \tiny{\cite{Shahzad2023AVLipSyncLA, liang2024speechforensics, Feng2023SelfSupervisedVF, oorloff2024avff, smeu2024circumventing}} & Audio+Video & Effective with limited fake data; sync-dependent, lacks adversarial robustness \\
\textbf{ZS Approaches} & \tiny{\cite{Cozzolino2022AudioVisualPD, li2024zero, reiss2023detecting, Jia2024CanCD, foteinopoulou2024hitchhiker}} & Aud+Vid+Txt & Adapts to unseen methods, minimal training; challenged by high-fidelity fakes, semantic reliance \\
\textbf{STF Analysis} & \tiny{\cite{Wang_2023_CVPR, Wang2023DynamicGL, zhou20253d, nguyen2025vulnerability, peng2024deepfakes}} & Spat+Temp+Freq & Captures subtle artifacts; high computational cost, latency concerns \\
\textbf{Exp Detection} & \tiny{\cite{li2024fakebench, peng2024deepfakes, Zhang2024CommonSR, foteinopoulou2024hitchhiker, Shao2023DetectingAG}} & Img+Txt+Vid & Enhances interpretability; limited reasoning depth, dataset-specific \\
\midrule
\multicolumn{4}{l}{\textbf{Datasets by Modality:}} \\
\textbf{Image+Text} & \multicolumn{3}{p{11cm}}{\tiny D3, CDDB-Hard, FakeClass, FakeClue, FakeQA, FF++, Celeb-DF, WildDeepfake, DGM4, StyleGAN2, Latent Diffusion, COCO, Flickr, SD2, SD3, SDXL, DALL-E, ProGAN, Twitter} \\
\textbf{Audio+Video} & \multicolumn{3}{p{11cm}}{\tiny FF++, DFDC, FakeAVCeleb, AVLips, DeepfakeTIMIT, LRS2, LRS3, KoDF, pDFDC, FaceForensics++, AV-Deepfake1M, LAV-DF, VidTIMIT} \\
\textbf{Aud+Vid+Txt} & \multicolumn{3}{p{11cm}}{\tiny FakeAVCeleb, DFDC, ASVS2015, ASVS2021LA, ASVS2021DF, MUSIC-21, DF-TIMIT, FakeOrReal, InTheWild, DefakeAVMiT, RAFV} \\
\textbf{Spat+Temp+Freq} & \multicolumn{3}{p{11cm}}{\tiny FF++, Celeb-DF, WildDeepfake, DFD, DFDCP, DF-v1.0, CDF1, CDF2, Celeb-DFv2, DFW} \\
\bottomrule
\end{tabular}
\scriptsize
\flushleft
Note: Abbreviations: VLM = Vision-Language Models, AV Sync = Audio-Visual Sync, MV Fusion = Multi-View Fusion, SS Learning = Self-Supervised Learning, ZS Approaches = Zero-Shot Approaches, STF Analysis = Spatial-Temporal-Frequency Analysis, Exp Detection = Explainable Detection; Aud+Vid+Txt = Audio + Video + Text, Spat+Temp+Freq = Spatial + Temporal + Frequency. Datasets are grouped by modality combinations, reflecting common evaluation contexts.
\end{table}
\endgroup

A systematic synthesis of recent advancements, as presented in Table~\ref{tab:multimodal_overview}, delineates seven paradigmatic categories of multi-modal detection, each leveraging distinct learning strategies and modality combinations. Vision-language models (VLM) harness supervised learning to align image-text representations (Img+Txt), with frameworks like, DE-FAKE \cite{sha2023fake}, MM-Det \cite{on-learning-multi-modal-forgery-representation-for-diffusion-generated-video-detection}, Prompt2Guard \cite{Laiti2024ConditionedPF}, AntifakePrompt \cite{Chang2023AntifakePromptPV}, and Bi-LORA \cite{keita2025bi, keita2024harnessing} optimizing pre-trained architectures (e.g., CLIP) for cross-modal feature extraction, achieving robust generalization across diffusion-based datasets. Audio-visual synchronization (AV Sync) employs supervised temporal modeling, exemplified by AMSDF \cite{wu2024audio}, LipFD \cite{liu2024lips}, AVSecure \cite{Guo2024AVSecureAA} and AV-MAE \cite{mo2024unveiling}, which integrate Aud+Vid signals via correlation analysis or watermarking to mitigate natural perturbations. Multi-view fusion (MV Fusion) consolidates audio, video, and text (Aud+Vid+Txt) through supervised transformer-based architectures, as in TMI-Former \cite{yoon2024triple}, FakeSTormer \cite{nguyen2025vulnerability}, and AVT²-DWF \cite{10609529}, enhancing cross-dataset transferability (e.g., DFDC, FakeAVCeleb) by modeling multi-dimensional feature interactions.

Complementary paradigms extend this methodological diversity. Spatial-temporal-frequency (STF) analysis, including SFDG \cite{Wang2023DynamicGL}, 3D ConvNet \cite{Wang_2023_CVPR}, and DBNet \cite{zhou20253d}, utilizes supervised dynamic graph or 3D-temporal learning to capture Spat+Temp+Freq inconsistencies, excelling at subtle artifact detection. Self-supervised learning (SS Learning) frameworks, such as SpeechForensics \cite{liang2024speechforensics}, Feng et al.’s anomaly detection \cite{Feng2023SelfSupervisedVF}, AVFF \cite{oorloff2024avff} and AV‑HuBERT \cite{smeu2024circumventing}, leverage Aud+Vid real-data distributions with contrastive or similarity-based objectives, offering efficacy in low-resource settings like FakeAVCeleb. Zero-shot (ZS) approaches, exemplified by CCFD \cite{li2024zero}, FACTOR \cite{reiss2023detecting}, and Jia et al.’s LLM integration \cite{Jia2024CanCD}, exploit pre-trained models or intrinsic Aud+Vid+Txt consistency for adaptability to novel manipulations sans retraining. Explainable detection (Exp Detection) methods, such as FakeBench \cite{li2024fakebench}, DD-VQA \cite{Zhang2024CommonSR}, and HAMMER \cite{Shao2023DetectingAG}, employ supervised or zero-shot reasoning over Img+Txt+Vid to elucidate forgery mechanisms, validated across FF++ and Celeb-DF benchmarks (Table~\ref{tab:multimodal_overview}).

These advancements signify a paradigm shift toward holistic detection, yet critical challenges temper their potential, as cataloged in Table~\ref{tab:multimodal_overview}. Modality alignment remains problematic under unsynchronized conditions \cite{Yang2023AVoiDDFAJ} or non-canonical perspectives \cite{kharel2023df, peng2024deepfakes}, despite robust performance against natural distortions \cite{Guo2024AVSecureAA, Cozzolino2022AudioVisualPD}. Adversarial robustness is underexplored, with notable exceptions like AVA-CL \cite{zhang2024joint} and MSOC \cite{lee2024multi} employing contrastive or one-class strategies to counter occlusions. Computational overhead restricts real-time applicability in resource-intensive frameworks \cite{Wang2023DynamicGL, kharel2023df}, while high-fidelity synthetics challenge ZS methods \cite{li2024zero}. Although generalizability is evident in cross-modal regularization \cite{zou2024cross}, transformer fusion \cite{koutlis2024dimodif}, and knowledge distillation \cite{Liu2024MCLMC}, linguistic diversity \cite{liu2024lips} and subtle manipulations \cite{Shao2023DetectingAG} expose persistent gaps.

This multi-modal transition reflects the escalating sophistication of synthetic media, necessitating a strategic research agenda. Enhancing adversarial resilience through targeted training \cite{nguyen2025vulnerability}, optimizing computational efficiency via lightweight designs \cite{wu2024audio}, and exploring nuanced forensic cues—e.g., gaze dynamics \cite{peng2024deepfakes} or micro-expressions—represent critical imperatives. Concurrently, bolstering interpretability \cite{Zhang2024CommonSR} and scalability \cite{reiss2023detecting} will bridge the gap between theoretical innovation and practical deployment, ensuring resilience against an evolving threat landscape. Table~\ref{tab:multimodal_overview} encapsulates this state-of-the-art synthesis, delineating mechanisms, modalities, and research frontiers in multi-modal deepfake detection.

\section{Deepfake Edited Regions Detection and localization}
\label{sec:localization}
Detecting and localizing edited regions in digital content is a critical task in deepfake analysis, necessitating methodologies that precisely delineate manipulated areas across images, videos, and audio. This section systematically reviews the technical approaches developed to address this challenge, tracing the progression from traditional forensic techniques to advanced deep learning frameworks, with an emphasis on the core strategies employed for identifying tampered regions.

Initial efforts leveraged traditional forensic techniques, focusing on statistical anomalies such as noise inconsistencies and compression artifacts. Methods like those in \cite{guillaro2023trufor, bai2024image, triaridis2024exploring, dagar2025noise} utilized noise fingerprints and spatial rich models to detect low-level tampering traces in static images, while \cite{cannas2024jpeg} explored counter-forensic impacts of neural compression using convolutional feature extraction. These approaches established a baseline but struggled with scalability against complex manipulations, driving the adoption of deep learning solutions.

Deep learning introduced supervised convolutional neural networks (CNNs) for localization, with hierarchical frameworks enhancing precision. \cite{guo2023hierarchical} employed multi-branch feature extractors and localization modules for fine-grained tamper detection, while \cite{guillaro2023trufor} integrated noise-sensitive fingerprints with RGB features to generate anomaly maps. Vision Transformers (ViTs) advanced this further: \cite{ma2023iml} combined windowed ViTs with multi-scale feature extraction, and \cite{su2024can} utilized sparse self-attention to target non-semantic artifacts. Lightweight strategies, such as \cite{guo2025lightweight}, adopted state space models for efficient multi-scale analysis, improving scalability.

Noise-guided methodologies became prominent for exposing subtle edits. \cite{zhu2024learning} fused denoising networks with cross-attention filters, while \cite{zhang2024samif} merged ViT and CNN branches to capture local noise cues for inpainting detection. Contrastive learning approaches, like \cite{bai2024image} and \cite{lou2025exploring}, applied multi-scale feature fusion and pixel-level contrast to isolate tampered regions, and \cite{tantaru2024weakly} explored weakly-supervised localization via patch-based scoring with an Xception backbone. Specialized techniques included \cite{lee2024localization}, using frequency-domain inter-intra similarity modules, and \cite{yao2025dense}, employing reverse edge-attention for inpainting boundary refinement.

Proactive forensics introduced watermarking strategies. \cite{Zhang2023EditGuardVI} and \cite{Zhang2024OmniGuardHM} embedded dual watermarks with invertible networks and adaptive transforms, while \cite{bai2025pim} mined inconsistencies through progressive feature refinement. \cite{han2024hdf} utilized a dual-stream architecture with coarse-to-fine localization, and \cite{Zhou2023PretrainingfreeIM} applied non-mutually exclusive contrastive learning to tackle data scarcity.

\begingroup
\scriptsize 
\setlength{\tabcolsep}{3pt} 
\renewcommand{\arraystretch}{1.3} 

\begin{table}[ht]
\centering
\caption{Overview of Edited Regions Detection and Localization Mechanisms} \label{tab:deepfake_localization}
\begin{tabular}{@{}p{3.5cm}p{2.5cm}p{1.4cm}p{7.0cm}@{}} 
\toprule
\textbf{Category} & \textbf{References} & \textbf{Modalities} & \textbf{Strengths and Challenges} \\
\midrule
\textbf{Traditional Forensic} & \tiny{\cite{guillaro2023trufor, bai2024image, triaridis2024exploring, dagar2025noise, cannas2024jpeg}} & Image & Detects statistical anomalies with noise and compression features; limited scalability to generative forgeries \\
\textbf{Uni-modal Image DL} & \tiny{\cite{guo2023hierarchical, ma2023iml, su2024can, guo2025lightweight, zhu2024learning, zhang2024samif, lou2025exploring, tantaru2024weakly, lee2024localization, yao2025dense, Zhou2023PretrainingfreeIM, DeCLIP}} & Image & Enables precise localization via hierarchical, noise-guided, and contrastive techniques; faces high computational demands and poor generalization to novel manipulations \\
\textbf{Proactive Image DL} & \tiny{\cite{Zhang2023EditGuardVI, Zhang2024OmniGuardHM, bai2025pim, han2024hdf}} & Image & Provides robust preemptive detection via watermarking; constrained by sensitivity to degradation and training data variability \\
\textbf{Multi-modal Image DL} & \tiny{\cite{wang2022objectformer, xu2024fakeshield, Li2024NoiseAssistedPL, luo2025toward, huang2024ffaa, Qu2024TowardsMI, zhu2024mesoscopic}} & Img+Txt & Leverages diverse cues for enhanced localization; challenged by data quality issues and increased inference complexity \\
\textbf{Uni-modal Video DL} & \tiny{\cite{zhao2023istvt, lin2024spatio, Hu2024DelocateDA, Xu2025LocalizationAD, lai2023detect, saha2023undercover, Shuai2023LocateAV}} & Video & Captures temporal artifacts with spatial-temporal models; limited by short sequence processing and sparse annotations \\
\textbf{Multi-modal Video DL} & \tiny{\cite{10609529, Jung2024WWWWW, Bharadwaj2024VANEBenchVA, cai2023glitch, cai20241m, cai2024av, Katamneni2024ContextualCA, ccma-ijcb, Zhang2024V2AMarkVD, zhang2023ummaformer}} & Vid+Aud & Achieves strong synergy across audio-visual modalities; hindered by alignment difficulties and focus on facial regions \\
\textbf{Audio DL} & \tiny{\cite{10890470, wu2024coarse, luong2025llamapartialspoof, roman2024proactive}} & Audio & Offers precise temporal forgery detection; restricted by limited adaptability and processing overhead \\
\textbf{Hybrid DL} & \tiny{\cite{mehta2025hfmf, miao2023multi}} & Img+Vid & Integrates flexible features across static and frame analysis; confined to specific domains with reduced scalability \\
\midrule
\multicolumn{4}{l}{\textbf{Datasets by Modality:}} \\
\textbf{Image} & \multicolumn{3}{p{10cm}}{\tiny CASIA, NIST16, Columbia, Coverage, HiFi-IFDL, CelebA, FFHQ, COCO, Places365, Dolos, WildRF, CollabDif, DMID, IID-74K, DEFACTO} \\
\textbf{Image+Text} & \multicolumn{3}{p{10cm}}{\tiny CASIA, NIST16, Columbia, Coverage, MMTD, RTM, FF++, Multi-attack, FFA-VQA} \\
\textbf{Video} & \multicolumn{3}{p{10cm}}{\tiny FF++, Celeb-DF, DFDC, GRIP, VideoSham, HTVD, FaceForensics++, DFD, FMLD, DeepfakeTIMIT, SORA, Vimeo-90K, Davis} \\
\textbf{Video+Audio} & \multicolumn{3}{p{10cm}}{\tiny DFDC, FakeAVCeleb, LAV-DF, AV-Deepfake1M, DeepfakeTIMIT, ForgeryNet, Psynd, TVIL} \\
\textbf{Audio} & \multicolumn{3}{p{10cm}}{\tiny ADD2023, PartialSpoof, LAV-DF, ASVspoof, VoxPopuli, LibriSpeech, Expresso, Half-truth} \\
\bottomrule
\end{tabular}
\scriptsize
\flushleft
Note: Abbreviations: DL = Deep Learning, Img = Image, Vid = Video, Aud = Audio, Txt = Text. References correspond to mechanisms in Section~\ref{sec:localization}. Datasets reflect evaluation contexts across modalities.
\end{table}
\endgroup

Multi-modal image-based methods integrated diverse cues for enhanced localization. \cite{wang2022objectformer} fused RGB and high-frequency features with object prototypes, \cite{xu2024fakeshield} employed domain-guided visual-textual analysis, and \cite{Li2024NoiseAssistedPL} adapted CLIP with noise-assisted prompts. \cite{luo2025toward} targeted text manipulation with transformed domain features, \cite{triaridis2024exploring} combined forensic filters via early fusion, and \cite{dagar2025noise} merged noise and spatial features hierarchically. \cite{huang2024ffaa} introduced a multimodal VQA framework, \cite{Qu2024TowardsMI} used adaptive perception with auto-annotation, and \cite{zhu2024mesoscopic} orchestrated mesoscopic features with a CNN-Transformer hybrid.

Video-based localization extended these principles to temporal analysis. \cite{zhao2023istvt} employed spatio-temporal transformers, \cite{lin2024spatio} applied co-attention fusion across frame streams, and \cite{zhang2023ummaformer} integrated temporal feature attention. \cite{Hu2024DelocateDA} utilized masked autoencoders with meta-learning, \cite{Xu2025LocalizationAD} leveraged synthetic self-blending, and \cite{lai2023detect} adapted SAM with multiscale adapters. \cite{saha2023undercover} combined ViT and timeseries transformers, and \cite{Shuai2023LocateAV} fused dual-stream modalities for verification.

Multi-modal video approaches incorporated audio-visual integration. \cite{10609529} utilized transformer-based dynamic fusion, \cite{Jung2024WWWWW} targeted clip-level localization, and \cite{Bharadwaj2024VANEBenchVA} assessed anomalies via VQA. \cite{cai2023glitch} combined multiscale ViTs, \cite{cai20241m, cai2024av} fused cross-modal interactions, and \cite{Katamneni2024ContextualCA, ccma-ijcb} applied contextual attention with recurrent units. \cite{Zhang2024V2AMarkVD} embedded watermarks for dual-modal tamper detection.

Audio-based methods addressed temporal forgery detection. \cite{10890470} employed adversarial training with Wav2Vec, \cite{wu2024coarse} refined proposals with coarse-to-fine learning, and \cite{roman2024proactive} embedded imperceptible watermarks. \cite{luong2025llamapartialspoof} assessed countermeasures using multi-resolution feature extraction. Hybrid approaches included \cite{miao2023multi}, localizing video frame edits with spectral analysis, \cite{mehta2025hfmf}, fusing hierarchical features for image deepfakes, and \cite{DeCLIP}, decoding CLIP embeddings for tampering detection.

Across the spectrum of methodologies reviewed—from traditional forensic techniques to advanced deep learning frameworks spanning static images, videos, and audio—the proposed approaches demonstrate notable strengths in detecting and localizing edited regions within digital content. Traditional forensic methods excel at identifying statistical anomalies, while deep learning variants leverage hierarchical feature extraction, noise-guided analysis, watermarking, and multi-modal cue integration to achieve precise tamper delineation. These techniques collectively enhance the ability to pinpoint manipulations with increasing sophistication, adapting to the evolving complexity of deepfake technologies. However, a pervasive limitation emerges in their collective evaluation: an almost universal absence of rigorous testing against adversarial robustness, as summarized in Table~\ref{tab:deepfake_localization}. Despite their efficacy under controlled conditions, the susceptibility of these methods to adversarial perturbations—such as subtle input modifications designed to evade detection—remains largely unaddressed, as evidenced by the consistent omission of such assessments across the referenced works. This gap is particularly critical in deepfake analysis, where adversarial attacks could exploit vulnerabilities in feature extraction or model decision boundaries, rendering localization unreliable in real-world scenarios. The absence of adversarial evaluation underscores a significant challenge to the practical deployment of these methods, highlighting the urgent need for future research to incorporate robust adversarial testing to ensure resilience against malicious countermeasures, thereby strengthening the integrity of deepfake detection and localization frameworks.

\section{Open Challenges and Future Directions}
Despite notable advancements in deepfake detection, the field continues to grapple with several unresolved challenges that impede the creation of robust and reliable systems, particularly as generative technologies and adversarial threats evolve rapidly. These challenges encompass generalizability to emerging generative models, robustness against adversarial and natural perturbations, computational efficiency for real-time deployment, effective multi-modal integration, precise localization of subtle manipulations, and the availability of diverse datasets. Among these, this review identifies the evaluation of adversarial robustness as a critical yet underexplored gap within the existing literature. This section first outlines these broad challenges, then delves into the specific issue of adversarial robustness—examining key attack strategies, their targeted modalities, and the detection models they undermine—and concludes by proposing future directions to address these limitations, emphasizing adaptive frameworks and cross-modal knowledge sharing.

\subsubsection*{Broad Challenges in Deepfake Detection}

The relentless advancement of generative models, such as diffusion-based architectures and transformer-based language models, poses significant hurdles for detection systems, which often struggle to generalize beyond the specific artifact patterns they were trained to recognize. This challenge is compounded by the need to maintain performance under natural perturbations—such as compression, environmental noise, lighting variations, or linguistic noise—that can obscure manipulation cues across image, video, audio, and text modalities. Computational efficiency remains a critical constraint, particularly for real-time applications where resource limitations demand lightweight yet effective models. The integration of multi-modal data—spanning image, video, audio, and text—introduces additional complexity, as aligning features across disparate domains requires overcoming modality-specific noise and inconsistencies. Precise localization of subtle manipulations, such as minor facial distortions, imperceptible audio splicing, or contextually inconsistent text, is hindered by insufficient labeled data, while the scarcity of diverse datasets capturing a wide range of manipulation techniques and real-world conditions restricts model training and evaluation.

These challenges vary across media types. For image-based deepfakes, detection systems must address high-resolution details and compression artifacts. Video-based systems face temporal inconsistencies and motion-based anomalies. Audio deepfakes present unique difficulties, such as identifying unnatural speech patterns or splicing in noisy environments, while text-based deepfakes require detecting subtle linguistic manipulations, like contextually inappropriate phrasing or adversarial perturbations. Recognizing these modality-specific issues is essential for developing comprehensive detection frameworks capable of addressing the multifaceted nature of synthetic media.

\subsubsection*{Adversarial Robustness: A Critical Yet Missing Evaluation}

Adversarial robustness is a cornerstone of reliable deepfake detection, yet many proposed models lack thorough assessments against adversarial perturbations, leaving them vulnerable to an increasingly sophisticated array of threats. This deficiency exposes systems to attacks that exploit weaknesses across modalities—image, video, audio, and text—compromising a diverse spectrum of detection frameworks, from convolutional neural networks (CNNs) and frequency-domain analyzers to watermark-based and transformer-based systems. The following adversarial attack strategies, drawn from recent literature, highlight these vulnerabilities:

\begin{itemize}
    \item \textbf{Image and Video Attacks}: 
        \begin{itemize}
            \item \textit{Pixel-Space Attacks}: Subtle alterations, such as blur, noise, or exposure adjustments \cite{hou2023evading, yang2024adversarial}, and natural shadow overlays \cite{liu2024advshadow}, evade spatial detectors (e.g., ResNet50, EfficientNet-b4) by aligning statistical distributions or masking differences \cite{Chen2023AdvINN}.
            \item \textit{Black-Box Perturbations}: Attacks targeting salient facial regions using Natural Evolutionary Strategies (NES) \cite{gowrisankar2024adversarial, tsigos2024towards}, super-resolution techniques \cite{coccomini2023adversarial}, or diffusion-based purification \cite{saberi2023robustness} challenge models like MesoNet, XceptionNet, and Swin-Small, achieving imperceptible yet potent degradation.
            \item \textit{Frequency-Domain Attacks}: Spectral manipulations \cite{abdullah2024analysis}, 2D convolutional filters \cite{galdi20242d}, and frequency-based Bayesian perturbations \cite{Diao2024VulnerabilitiesIA} degrade detectors like DCT and FrequencyForensics, exploiting spectral inconsistencies with high transferability.
            \item \textit{Latent-Space Attacks}: Perturbations in generative model representations, such as StyleGAN2’s latent space \cite{meng2024ava}, customized Stable Diffusion outputs \cite{abdullah2024analysis}, or diffusion-based latent optimization \cite{zhou2024stealthdiffusion, Sun_2024_CVPR, guo2024efficient, liu2023diffprotect}, bypass DNN-based and commercial detectors (e.g., Baidu, Tencent) with remarkable success rates.
            \item \textit{Backdoor Attacks}: Poisoned training data with embedded triggers \cite{kassis2024unlocking, liu2024robust} or diffusion process manipulations \cite{kassis2024unlocking} compromise models like WideResNet and DeiT-S, activating malicious behavior during inference.
            \item \textit{Watermarking Attacks}: Diffusion purification \cite{saberi2023robustness}, universal spectral-domain attacks \cite{kassis2025unmarker}, and adversarial watermark fine-tuning \cite{wu2024watermarks} undermine watermark-based detectors (e.g., StegaStamp, RivaGAN), significantly reducing detection efficacy.
        \end{itemize}
    \item \textbf{Audio Attacks}: Adversarial noise injection or splicing \cite{liu2024robust}, white-box attacks like FGSM and PGD \cite{kawa2022defense}, and GAN-based transferable attacks \cite{farooq2025transferable} exploit temporal inconsistencies, targeting neural network detectors (e.g., LCNN, RawNet3) and end-to-end models (e.g., RawNet2, Res-TSSDNet), with detection accuracies dropping sharply (e.g., 98\% to 26\%).
    \item \textbf{Text Attacks}: Adversarial perturbations, such as synonym substitution or grammatical alterations \cite{liu2024robust}, and low-cost attacks like decoding strategy shifts and DFTFooler \cite{pu2023deepfake}, mislead text-based detectors (e.g., GROVER, BERT-Defense) by preserving semantic meaning while altering syntactic or statistical features, achieving high evasion rates (up to 91.3\%).
\end{itemize}

These strategies collectively underscore systemic frailties across detection paradigms, emphasizing the urgent need for comprehensive robustness evaluations that encompass the full spectrum of modalities and adversarial threats identified in current research.

\subsubsection*{Future Directions: Adaptive Frameworks and Cross-Modal Knowledge Sharing}

To address these challenges, particularly the critical gap in adversarial robustness, future research must prioritize the development of adaptive detection frameworks that leverage cross-modal knowledge sharing to enhance resilience across image, video, audio, and text modalities. Drawing on recent advancements, such frameworks could enable detectors to dynamically adjust their focus based on input characteristics and task demands, improving their ability to identify subtle manipulations and withstand diverse adversarial perturbations.

A key strategy involves designing modality-agnostic architectures that integrate and transfer knowledge across media types. For instance, image-based detectors could incorporate audio cues to detect lip-sync inconsistencies, while text-based systems might verify narrative consistency with visual elements, mitigating modality-specific attacks by diversifying feature representations \cite{wang2023connecting, zhang2024extending, nguyen2025capsfake}. Expert-driven architectures, with specialized modules targeting distinct manipulation types—such as facial distortions in video, voice cloning in audio, or syntactic anomalies in text—could be dynamically weighted based on input reliability, enhancing sensitivity to cross-modal anomalies \cite{NEURIPS2024_009729d2}.

To counter the rapid evolution of deepfake technologies, parameter-efficient adaptation techniques, such as lightweight projectors or adapters, could facilitate rapid recalibration to emerging manipulation methods without extensive retraining. Unsupervised or semi-supervised learning approaches could reduce reliance on scarce labeled datasets, leveraging abundant unlabeled data to improve robustness. Explainable AI techniques could enhance localization capabilities, providing interpretable insights into detected manipulations and informing targeted countermeasures \cite{pinhasov2024xai}.

Standardized benchmarks encompassing a broad spectrum of adversarial scenarios—spanning pixel-space, frequency-domain, latent-space, backdoor, and watermarking attacks across all modalities—are essential for consistent and rigorous robustness assessments. Datasets enriched with adversarial perturbations, reflecting real-world conditions, will further support model generalization. Lightweight models balancing computational efficiency with resilience will ensure practical deployment in resource-constrained environments.

Finally, the societal implications of deepfake detection failures—such as misinformation, fraud, or privacy breaches—underscore the need for robust systems. Future efforts should integrate ethical considerations, developing transparent and accountable frameworks to safeguard public trust and security.

In summary, while deepfake detection faces a multifaceted array of challenges, the insufficient evaluation of adversarial robustness remains a pivotal shortfall. By adopting adaptive frameworks with cross-modal knowledge sharing, embedding adversarial training, and establishing standardized benchmarks, the field can advance toward reliable, scalable detection systems capable of countering the dynamic and escalating threats posed by synthetic media.

\section{Conclusion}
The escalating sophistication of Generative Artificial Intelligence has amplified the deepfake threat across image, video, audio, and text modalities, challenging the integrity of digital systems and societal trust. This systematic review has elucidated the strengths and limitations of contemporary deepfake detection methodologies, spanning uni-modal and multi-modal frameworks adept at identifying fully synthetic media and localizing subtle manipulations within authentic content. While these approaches demonstrate commendable precision in controlled settings, their vulnerability to adversarial perturbations and limited generalizability to emerging generative techniques underscore a critical gap in real-world applicability.

Our analysis highlights the urgent need for robust, adaptable detection systems capable of withstanding the evolving landscape of synthetic media threats. By prioritizing reproducibility through a curated repository of open-source implementations, this study fosters transparency and enables rigorous validation of current methods. The integration of multi-modal cues emerges as a promising avenue, yet the pervasive shortfall in adversarial robustness demands innovative solutions beyond traditional paradigms. Future advancements must focus on scalable, modality-agnostic architectures that enhance resilience, alongside standardized evaluation protocols to ensure consistent performance across diverse scenarios.

In synthesizing these insights, this review not only consolidates the current state of deepfake detection but also delineates a strategic path forward. The development of trustworthy systems—capable of mitigating misinformation, safeguarding privacy, and maintaining digital security—hinges on addressing these identified challenges with a concerted emphasis on robustness and adaptability. This study thus provides a foundational framework for advancing next-generation detection capabilities, poised to meet the complexities of an increasingly synthetic digital era.

\bibliographystyle{IEEEtran}
\bibliography{references}

\end{document}